\documentclass[preprint,showkeys,showpacs]{revtex4}
\usepackage{amssymb}
\usepackage{amsmath}
\usepackage{graphicx}
\renewcommand{\figurename}{Fig.}

\begin{document} 
\title{Neutron skin and halo in medium and heavy nuclei within the extended Thomas-Fermi theory}
\author{S.V. Lukyanov and A.I. Sanzhur}
\affiliation{Institute for Nuclear Research, 03680 Kyiv, Ukraine}

\begin{abstract}
The neutron skin and halo distributions in medium and heavy nuclei are calculated within the extended 
Thomas-Fermi approximation. Calculations are carried out for the effective Skyrme-like forces using 
the direct variational method. The analytical expression for the isovector shift of the rms radii 
$\Delta r_{np}$ as a sum of skin- and halo-like terms is obtained. The contribution of halo and skin 
terms to $\Delta r_{np}$ are found to be approximately equal.
\end{abstract}

\maketitle

\section{Introduction}
For the analysis of the experimental data the simplest two-parameter Fermi (2pF) distribution of 
normalized nucleon densities is often used:
\begin{equation}\label{2pf}
F_q(r)= \left[1+\exp\left(\frac{r-R_q}{a_q}\right) \right]^{-1},
\end{equation}
where $R_q$ is the half-density radius and $a_q$ is the diffuseness parameter of the distribution. 
Here $q=n$ is for neutron and $q=p$ for proton distributions.
There are two opposite pictures in description of a two-component finite Fermi system
with 2pF. The first one is the so-called "neutron skin-type" distribution having the neutron
half-density radius larger than the proton half-density radius, $R_n>R_p$, and equal diffuseness 
parameters $a_n=a_p$. The second one is the "neutron halo-type" distribution having $R_n=R_p$ and 
$a_n>a_p$.  A mixture of the neutron "skin-type" and "halo-type" distributions having $R_n>R_p$ 
and $a_n>a_p$ is also possible. It is found that the experimental data can be reproduced by a variety 
of these 2pF distributions with different values of $R_n-R_p$ and $a_n-a_p$ \cite{waviroce10}.  
In that sense there is an ambiguity in theoretical description of the experimental data. 

In the present paper we study this ambiguity within the extended Thomas-Fermi approximation (ETFA) 
based on the direct variational method. The nucleon densities $\rho_q(r)$ are generated 
by the profile functions which are obtained from the requirement that the energy of the nucleus should 
be stationary with respect to variations of these profiles. The profile functions fulfill 
the leptodermous conditions and take into account some asymmetry of the nucleon distributions like 
in real nuclei.

\section{The model}

In general, the total energy of a nucleus includes the kinetic and potential energies and is given by
\cite{kirz67,brguha85,book}
\begin{equation}
E_\mathrm{tot}=\int d\mathbf{r}\ \left[\epsilon_\mathrm{kin}(\mathbf{r})+\epsilon_\mathrm{pot}(\mathbf{r})\right],
\label{etot}
\end{equation}
where the potential energy density, $\epsilon_\mathrm{pot}(\mathbf{r})$, includes the NN-interaction,
$\epsilon_\mathrm{NN}(\mathbf{r})$, the spin-orbit part of the NN-interaction, $\epsilon_\mathrm{SO}(\mathbf{r})$,
and the Coulomb energy density, $\epsilon_\mathrm{C}(\mathbf{r})$:
\begin{equation}
\epsilon_\mathrm{pot}(\mathbf{r})=\epsilon_\mathrm{NN}(\mathbf{r})+\epsilon_\mathrm{SO}(\mathbf{r})
+\epsilon_\mathrm{C}(\mathbf{r}).
\label{edenstot}
\end{equation}

In the framework of ETFA, the kinetic energy density is given by the sum of the neutron and proton 
contributions
\cite{brguha85}
\begin{equation}
\epsilon_\mathrm{kin}(\mathbf{r})=\epsilon_\mathrm{kin,n}(\mathbf{r})+\epsilon_\mathrm{kin,p}(\mathbf{r}),
\label{ekinsum}
\end{equation}
where
\begin{equation}
\epsilon_\mathrm{kin,q}(\mathbf{r})
={\frac{\hbar^{2}}{2m}}\left[ {\frac{3}{5}}\,(3\,\pi ^{2})^{2/3}\,\rho _{q}^{5/3}
+\eta\frac{1}{36}{\frac{(\mathbf{\nabla }\rho _{q})^{2}}{\rho _{q}}}+{\frac{1}{3}}\,\nabla^{2}\rho _{q}\right].
\label{eqkin}
\end{equation}
Here $m$ is the bare nucleon mass.
The semiclassical consideration gives the value of the parameter $\eta=1$ in Eq. (\ref{eqkin}) \cite{kirz67,brguha85}.
In the asymptotic limit $r\rightarrow\infty$, the semiclassical particle density $\rho_q$
with $\eta=1$ falls off to zero significantly faster than the one from the quantum-mechanical calculation, where
one has $\eta=4$. We will use both values of the parameter $\eta$ to study the neutron halo and
skin appearances and their superposition in nuclei.

For the effective nucleon-nucleon interaction we will use the Skyrme force \cite{brguha85}
\begin{eqnarray}
\epsilon_\mathrm{NN}(\mathbf{r})&=&
\frac{t_0}{2} \left[\left(1+\frac{x_0}{2}\right)\rho^2-\left(x_0+\frac{1}{2}\right)\left(\rho_n^2+\rho_p^2\right)\right]
\nonumber \\
&+&\frac{t_3}{12}\rho^\nu \left[\left(1+\frac{x_3}{2}\right) \rho^2
-\left(x_3+\frac{1}{2}\right)\left(\rho_n^2+\rho_p^2\right)\right]
\nonumber \\
&+&\frac{1}{4}\left[t_1\left(1+\frac{x_1}{2}\right)+t_2\left(1+\frac{x_2}{2}\right)\right]\epsilon_{kin}\rho
\nonumber \\
&+&\frac{1}{4}\left[t_2\left(x_2+\frac{1}{2}\right)-t_1\left(x_1+\frac{1}{2}\right)\right]
\left(\epsilon_{kin,n} \rho_n +\epsilon_{kin,p} \rho_p\right)
\nonumber \\
&+&\frac{1}{16}\left[3t_1\left(1+\frac{x_1}{2}\right)-t_2\left(1+\frac{x_2}{2}\right)\right]\left(\mathbf{\nabla}\rho\right)^2
\nonumber \\
&-&\frac{1}{16}\left[3 t_1\left(x_1+\frac{1}{2}\right)+t_2\left(x_2+\frac{1}{2}\right)\right]
\left[\left(\mathbf{\nabla}\rho_n\right)^2+\left(\mathbf{\nabla}\rho_p\right)^2\right], \label{enn}
\end{eqnarray}
where $\rho=\rho_n+\rho_p$ is the total density of nucleons. The spin-orbit part of the Skyrme force is written as 
\begin{equation}
\epsilon_\mathrm{SO}(\mathbf{r})=-\frac{m^*(r)}{\hbar^2}\frac{W_0^2}{4}
\left[\rho_n \left(2 \mathbf{\nabla}\rho_n + \mathbf{\nabla}\rho_p \right)^2
+\rho_p \left( \mathbf{\nabla}\rho_n + 2 \mathbf{\nabla}\rho_p\right)^2 \right].
\end{equation}
Here $t_0$, $t_1$, $t_2$, $t_3$, $x_0$, $x_1$, $x_2$, $x_3$, $\nu$, $W_0$ are the parameters of the Skyrme 
interaction and the effective mass of a nucleon is
$$
m^*(r)=\frac{m}{1+\beta\rho(r)},
\quad \text{where} \quad
\beta=\frac{2m}{\hbar^2}\frac{1}{4}\left[\frac{1}{4}(3t_1+5t_2)+t_2x_2\right].
$$

At the moment there are plenty of parameter sets $t_i$, $x_i$, $\nu$ and $W_0$ for different
modifications of the Skyrme interaction. They are adjusted using the well-known properties of the nuclear 
matter and real nuclei and denoted by the letters as SI, SIII, SkM$^*$ and so on. Below we will use 
this kind of notifications.
The Coulomb energy density is taken in the well-known form \cite{book}
\begin{equation}
\epsilon_\mathrm{C}(\mathbf{r})=e^2\rho_p(\mathbf{r}) \frac{1}{2} \int d\mathbf{r'} \frac{\rho_p(\mathbf{r'})}{\vert\mathbf{r}
-\mathbf{r'}\vert}
-\frac{3}{4}e^2\left(\frac{3}{\pi}\right)^{1/3}\rho^{4/3}_p(\mathbf{r}).
\end{equation}

The unknown values of $\rho_q$ can be evaluated from the condition of equilibrium.
The corresponding condition implies that the total energy $E_\mathrm{tot}$ reaches a minimum value 
for a given number of neutrons, $N$, and protons, $Z$,
\begin{equation}
N=\int d\mathbf{r} \rho_n(\mathbf{r}), \quad Z=\int d\mathbf{r} \rho_p(\mathbf{r}).
\end{equation}

In general, one can use an arbitrary trial function as nucleon distribution. 
In any case the variational method with the effective nucleon-nucleon interaction (\ref{enn}) 
reduces the nucleon distribution to that like 2pF (\ref{2pf}). Following 
the direct variational method \cite{kosa08} the trial function for $\rho _{q}(\mathbf{r})$ is taken 
as a power of the Fermi function 
\begin{equation}
\rho_{q}(\mathbf{r})=\rho_{0,q}\left[1+\exp \left( \frac{r-R_{q}}{a_{q}}\right) \right]^{-\xi},
\label{rhoq}
\end{equation}
where $\rho_{0,q}$, $R_{q}$, $a_{q}$ and $\xi$ are the variational parameters. The variational 
parameter $\xi$ takes into account the asymmetry of the nucleon distributions around the nuclear surface.

\section{Numerical results}

The nuclear \textrm{rms} radius for nucleons is defined as
\begin{equation}
\sqrt{\left\langle r^2_q \right\rangle}
=\sqrt{\left.\int d\mathbf{r}\,r^2\,\rho_q(r) \right/ \int d\mathbf{r}\,\,\rho_q(r)}.
\label{rms}
\end{equation}
The difference between the neutron and proton root mean square radii gives the neutron-skin thickness
\begin{equation}
\Delta r_{np}=\sqrt{\left\langle r^2_n \right\rangle}-\sqrt{\left\langle r^2_p \right\rangle}.
\end{equation}

From the definition of trial functions it is seen that the value of $\Delta r_{np}$ can be described 
by the different radii $R_q$ (skin effect) and the  different diffusenesses $a_q$ (halo effect) of neutron 
and proton distributions, see also \cite{midolanare00}. Within the leptodermous approximation $a_q/R_q \ll 1$ 
one can separate the above contributions and the value of $\Delta r_{np}$ is written as
\begin{equation}
\Delta r_{np} = \Delta r_{np}^{skin}+\Delta r_{np}^{halo} + O \left(A^{-4/3}\right),
\label{drho1}
\end{equation}
where $\Delta r_{np}^{skin}$ and $\Delta r_{np}^{halo}$ are caused by the skin and halo effects, respectively.
The corresponding values are given by (see Appendix A)
\begin{eqnarray}
\Delta r_{np}^{skin}& \approx &\sqrt{\frac{3}{5}}\ \Delta_R \left\{1+\frac{7}{2}\left[\kappa_0^2(\xi) -2\kappa_1(\xi)\right]
\left( \frac{a}{R}\right)^2 \right. \nonumber \\
&  & - \left. \frac{1}{3}\left(75\kappa_0^3(\xi)-204\kappa_0(\xi)\kappa_1(\xi)+81\kappa_2(\xi)\right)\left(\frac{a}{R}\right)^3
\right\},
\label{drhoR}
\end{eqnarray}
and
\begin{eqnarray}
\Delta r_{np}^{halo}& \approx & \sqrt{\frac{3}{5}}\ \Delta_a \left\{ \kappa_0(\xi)
-7\left[\kappa_0^2(\xi)-2\kappa_1(\xi)\right] \frac{a}{R} \right. \nonumber \\
& & +\left. \frac{1}{2}\left(75\kappa_0^3(\xi)-204\kappa_0(\xi)\kappa_1(\xi)+81\kappa_2(\xi)\right)
\left(\frac{a}{R}\right)^2 \right\} ,
\label{drhoa}
\end{eqnarray}
where we have introduced the notations: $R=(R_n+R_p)/2$, $a=(a_n+a_p)/2$, $\Delta_R=R_n-R_p$, and $\Delta_a=a_n-a_p$, and
$\kappa_j(\xi)$ are the generalized Fermi integrals (see Appendix A).

\begin{figure}
\begin{center}
\includegraphics[scale=0.4,clip]{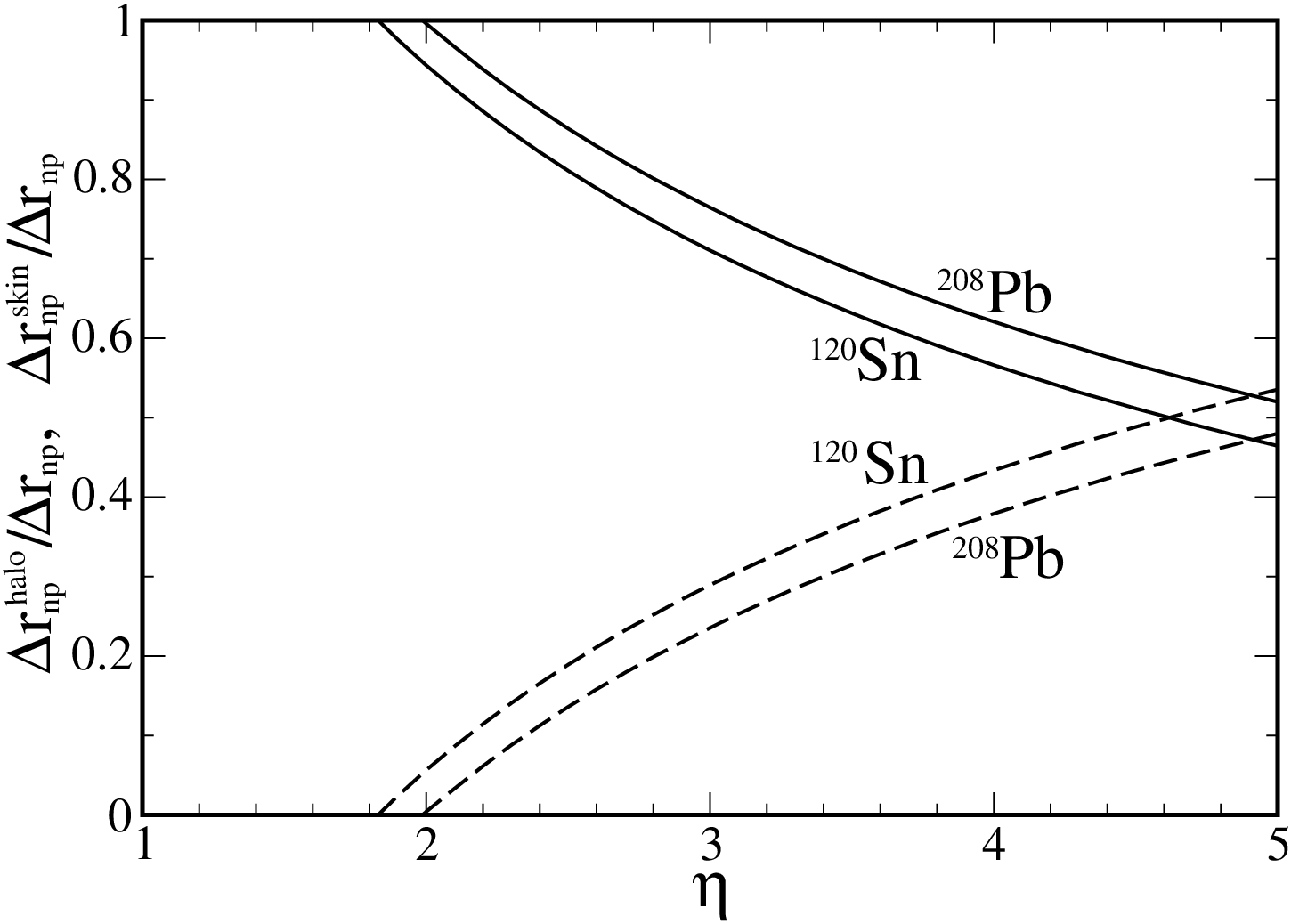}
\end{center}
\caption{The partial contribution to the isotopic shift of radii $\Delta r_{np}$ from the skin effect,
$\Delta r_{np}^\mathrm{skin}/\Delta r_{np}$ (solid lines) and the halo effect, $\Delta r_{np}^\mathrm{halo}/\Delta r_{np}$
(dashed lines) versus the parameter $\eta$. The calculations have been performed with SkM parametrization
for nuclei $^{120}$Sn and $^{208}$Pb.}
\label{fig1}
\end{figure}

\begin{figure}
\begin{center}
\includegraphics[scale=0.4,clip]{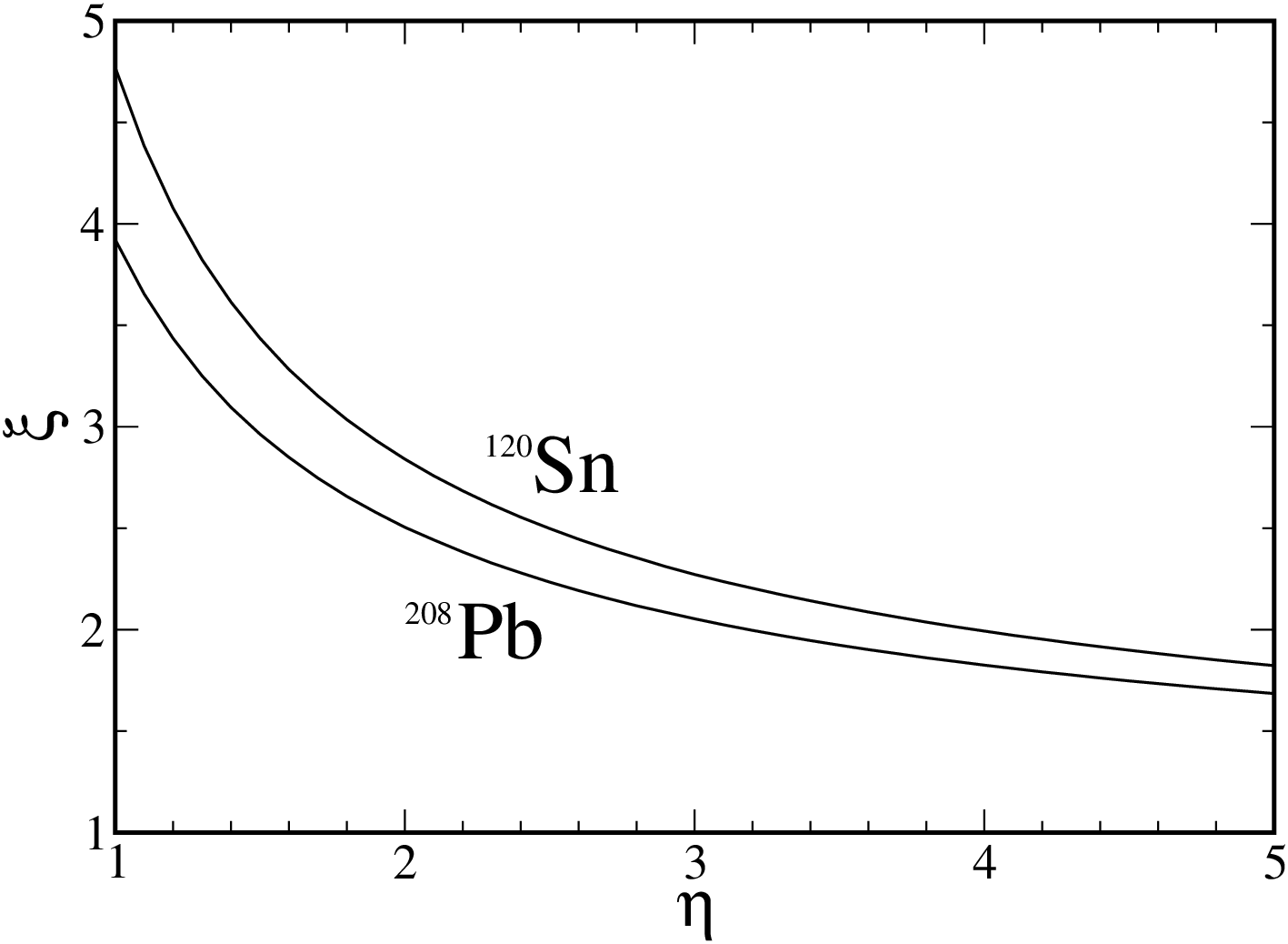}
\end{center}
\caption{Parameter $\xi$ versus $\eta$. The calculations have been performed for nuclei $^{120}$Sn and $^{208}$Pb
using SkM parametrization.}
\label{fig2}
\end{figure}

In \figurename~\ref{fig1} the partial contributions to the neutron-skin thickness from the skin effect,
$\Delta r_{np}^\mathrm{skin}/\Delta r_{np}$, and the halo effect, $\Delta r_{np}^\mathrm{halo}/\Delta r_{np}$,
are shown. One can see from \figurename~\ref{fig1} that the halo partial contribution increases and the skin 
partial contribution decreases with the increase of parameter $\eta$. Close to the semiclassical approach region, 
$1\leq\eta\leq 2$, the partial contributions from the halo effect are negative and therefore the partial 
contributions from the skin effect are greater than the unity. The reason for that is the high value of the
parameter $\xi$ for the asymmetric Fermi function, (see below \figurename~\ref{fig2}), so the densities fall 
off too quickly in the outer surface.
This means that the halo component is negative and reduces the difference between the neutron and proton root 
mean square radii. In the region $1\leq\eta\leq 2$ one has more preferably proton halo effect as
reported in \cite{waviroce10}. For the values of $\eta\sim 4-5$, where the tails of the densities are 
close to that obtained in the quantum mechanical calculations, the halo and skin components of 
$\Delta r_{np}$ are found to be approximately equal.

\begin{figure}
\begin{center}
\includegraphics[scale=0.4,clip]{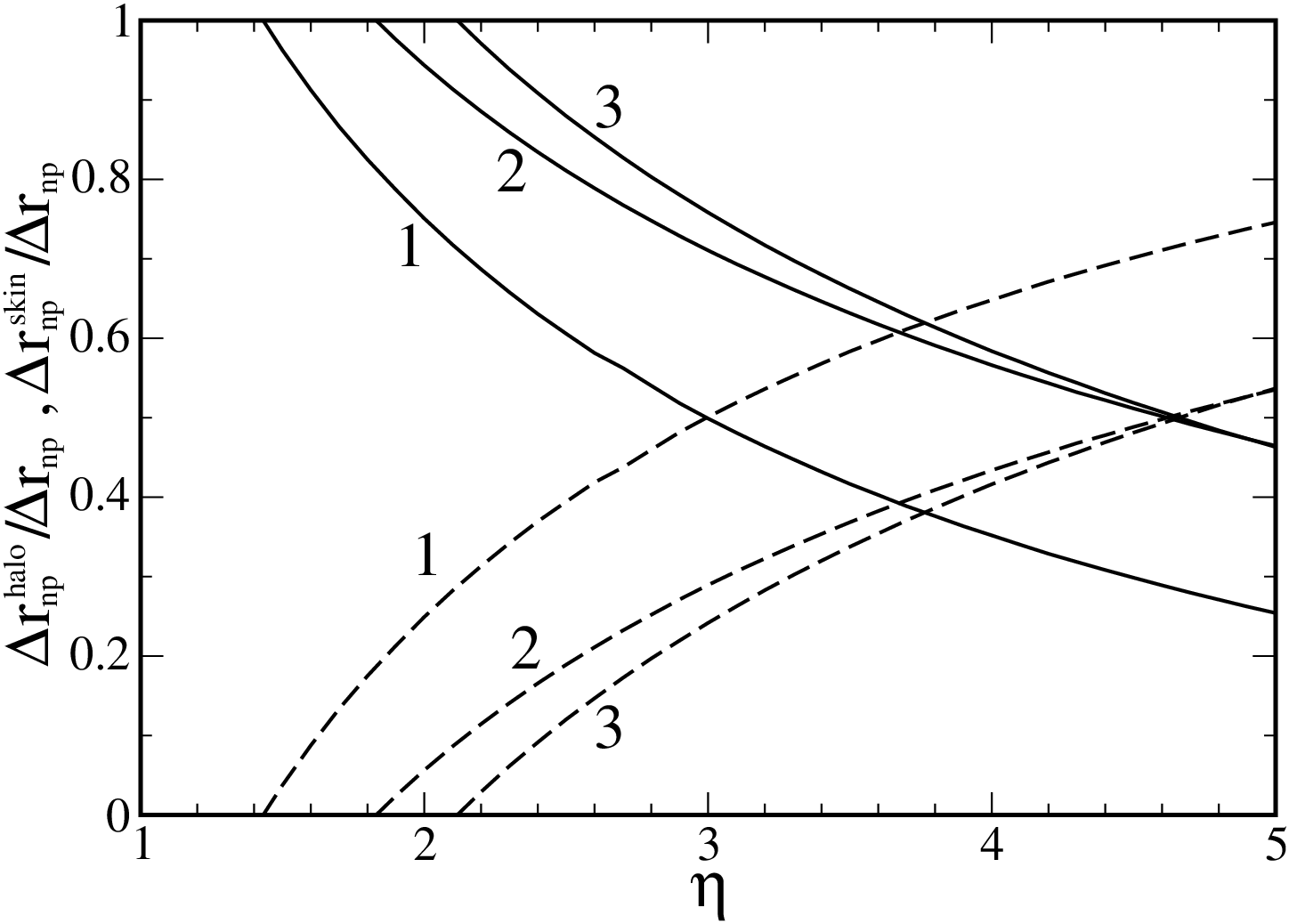}
\end{center}
\caption{The partial contribution to the isotopic shift of radii $\Delta r_{np}$ from the skin
effect, $\Delta r_{np}^\mathrm{skin}/\Delta r_{np}$ (solid lines) and the halo
effect, $\Delta r_{np}^\mathrm{halo}/\Delta r_{np}$ (dashed lines) versus the
parameter $\eta$. The calculations have been performed for $^{120}$Sn with the following Skyrme force
parametrizations: 1 - SIII, 2 - SkM, 3 - SLy230b.}
\label{fig3}
\end{figure}

The sensitivity of the partial contributions $\Delta r_{np}^\mathrm{skin}/\Delta r_{np}$ and
$\Delta r_{np}^\mathrm{halo}/\Delta r_{np}$ on the Skyrme force parametrization is shown in 
\figurename~\ref{fig3}. It is seen from \figurename~\ref{fig3} that the calculated values change 
of about 20\% for different parametrizations.

\begin{figure}
\begin{center}
\includegraphics[scale=0.4,clip]{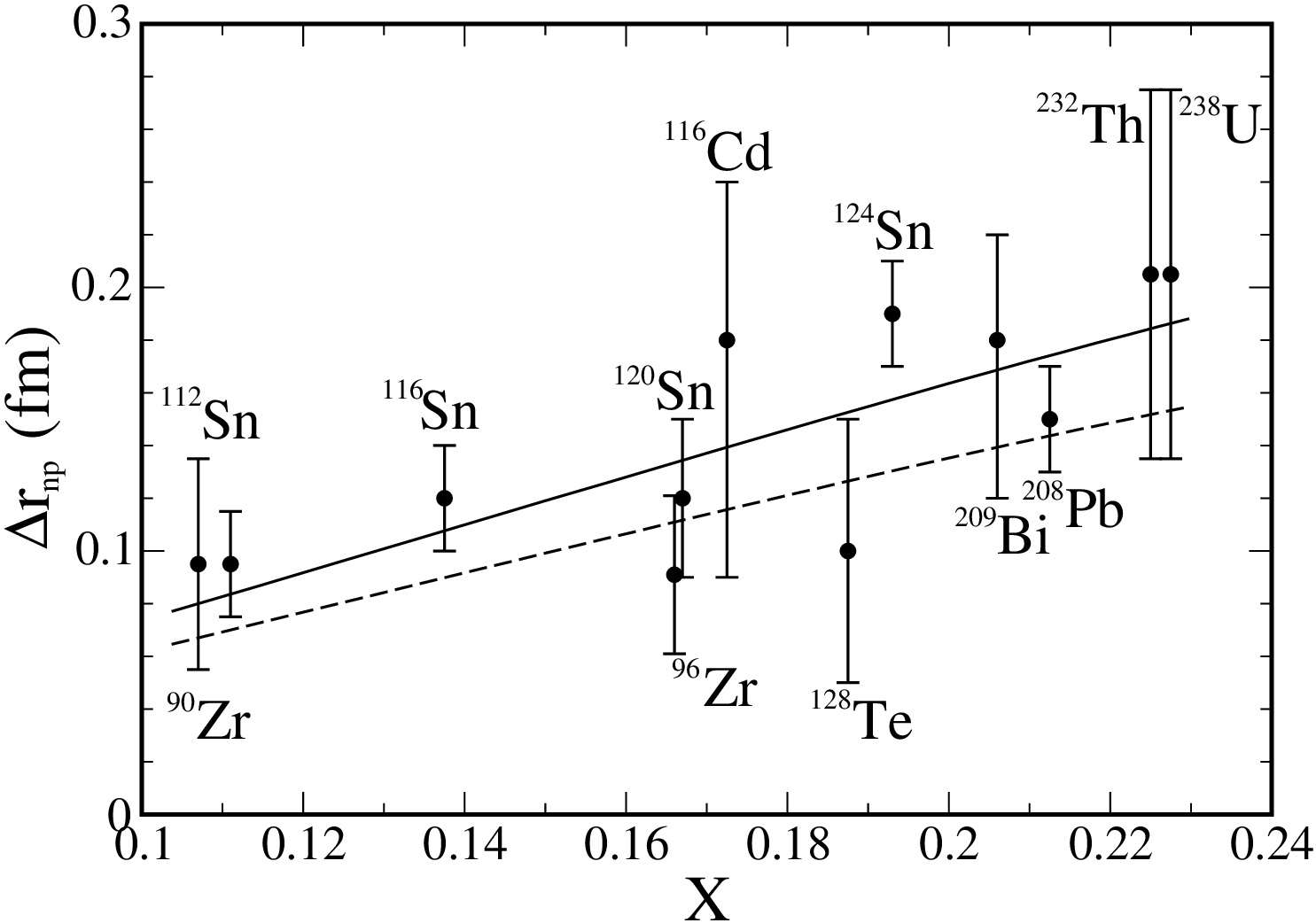}
\end{center}
\caption{Difference $\Delta r_{np}$ between the rms radii of the neutron and proton distributions as a function
of $X=(N-Z)/A$. The solid and dashed lines are calculations with $\eta=4$ and $2$ for SkM
parametrization of the Skyrme force. The experimental data are taken from Ref. \cite{trja01}.}
\label{fig4}
\end{figure}

The dependence of $\Delta r_{np}$ on the asymmetry parameter $X$ together with the experimental data 
\cite{trja01} are shown in \figurename~\ref{fig4}.
As it is seen from the figure the calculation with the "quantum mechanical" value $\eta=4$ (solid curve),
in general, agrees better with the experimental data than for $\eta=2$ (dashed curve). The obtained
dependencies are approximately linear for the considered values of isotopic asymmetry parameter $X$. 

\begin{figure}
\begin{center}
\includegraphics[scale=0.4,clip]{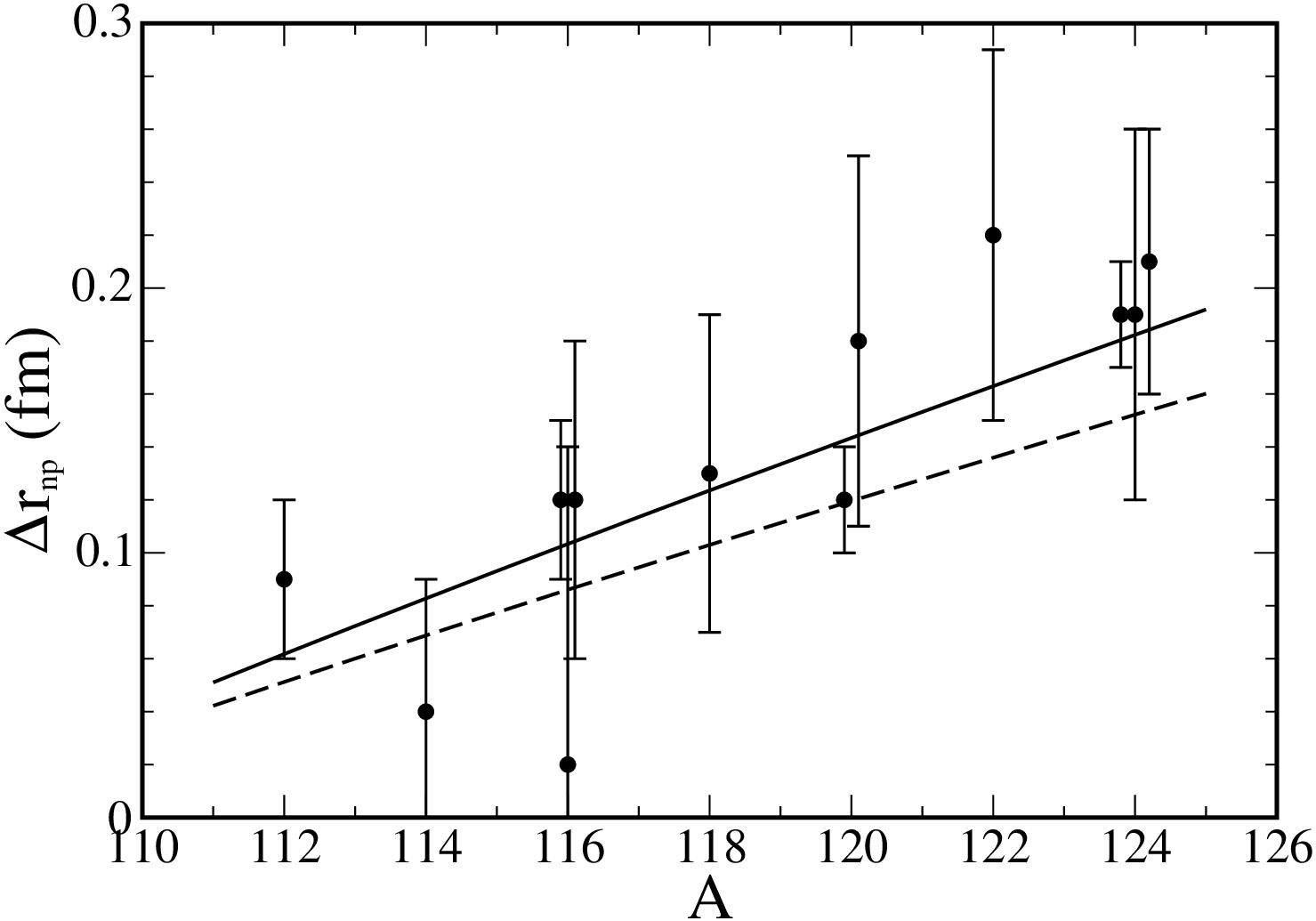}
\end{center}
\caption{Difference $\Delta r_{np}$ between the rms radii of the neutron and proton distributions as a function
of $A$ for Sn isotopes. The solid and dashed lines are calculations with $\eta=4$ and $2$ for SkM force.
The experimental data are taken from Ref. \cite{trja01}.}
\label{fig5}
\end{figure}

In \figurename~\ref{fig5} the analogous calculations for tin isotopes together with the experimental data are 
shown as a function of mass number. 
Similarly to the previous cases, the calculation with $\eta=4$ describes the experimental data better 
than the one with $\eta=2$. 

The neutron-to-proton density ratio in the peripheral region with respect to the bulk value is described
by the peripheral halo factor \cite{bug}. Following \cite{bug,trja01} we will also use the theoretical estimation
of the halo factor as
\begin{equation}
f_\mathrm{theor}^\mathrm{halo}(r)\approx \frac{\rho_n(r)}{\rho_p(r)}\frac{Z}{N}
\label{fhalo}
\end{equation}

\begin{figure}
\begin{center}
\includegraphics[scale=0.33,clip]{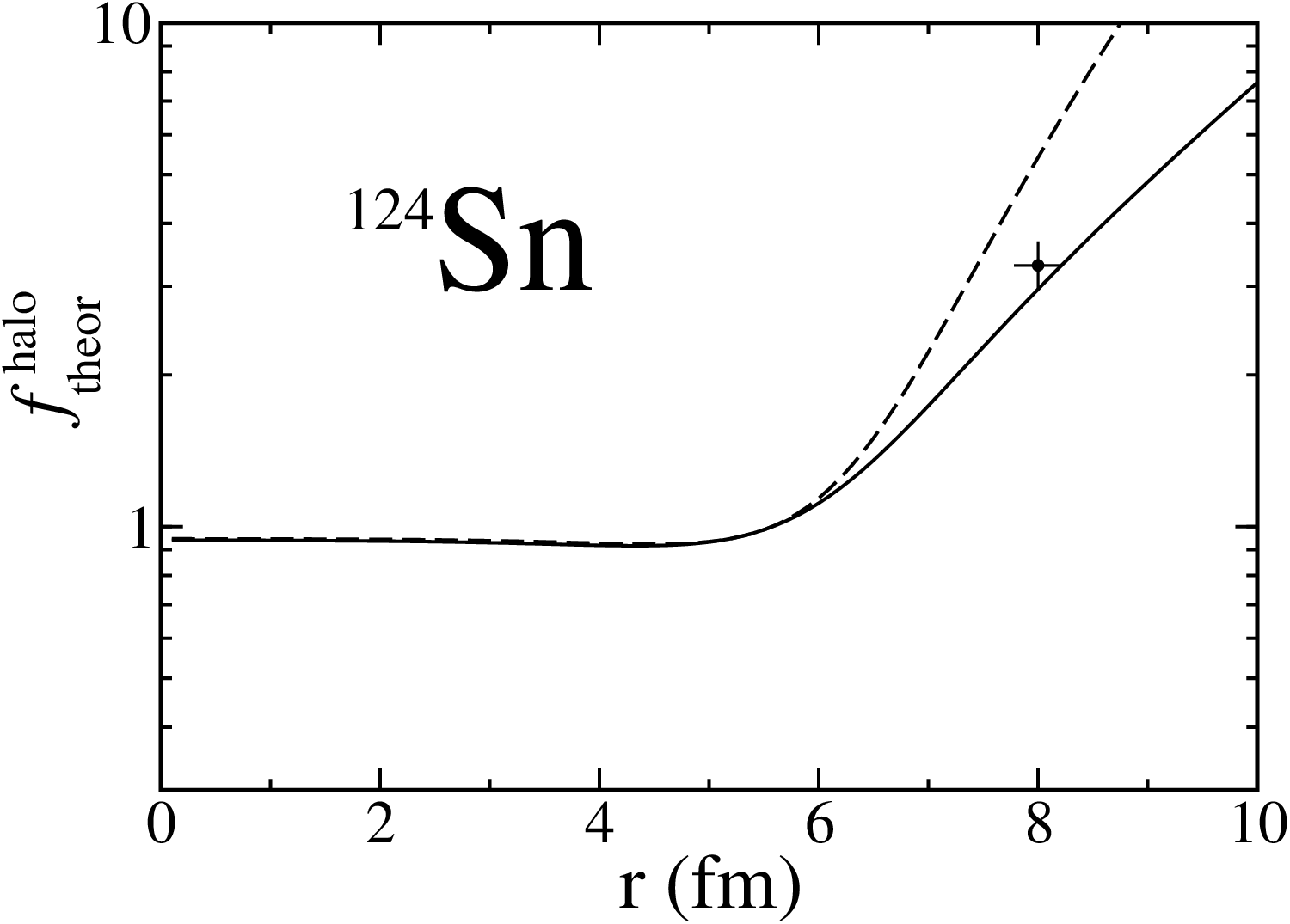}\includegraphics[scale=0.33,clip]{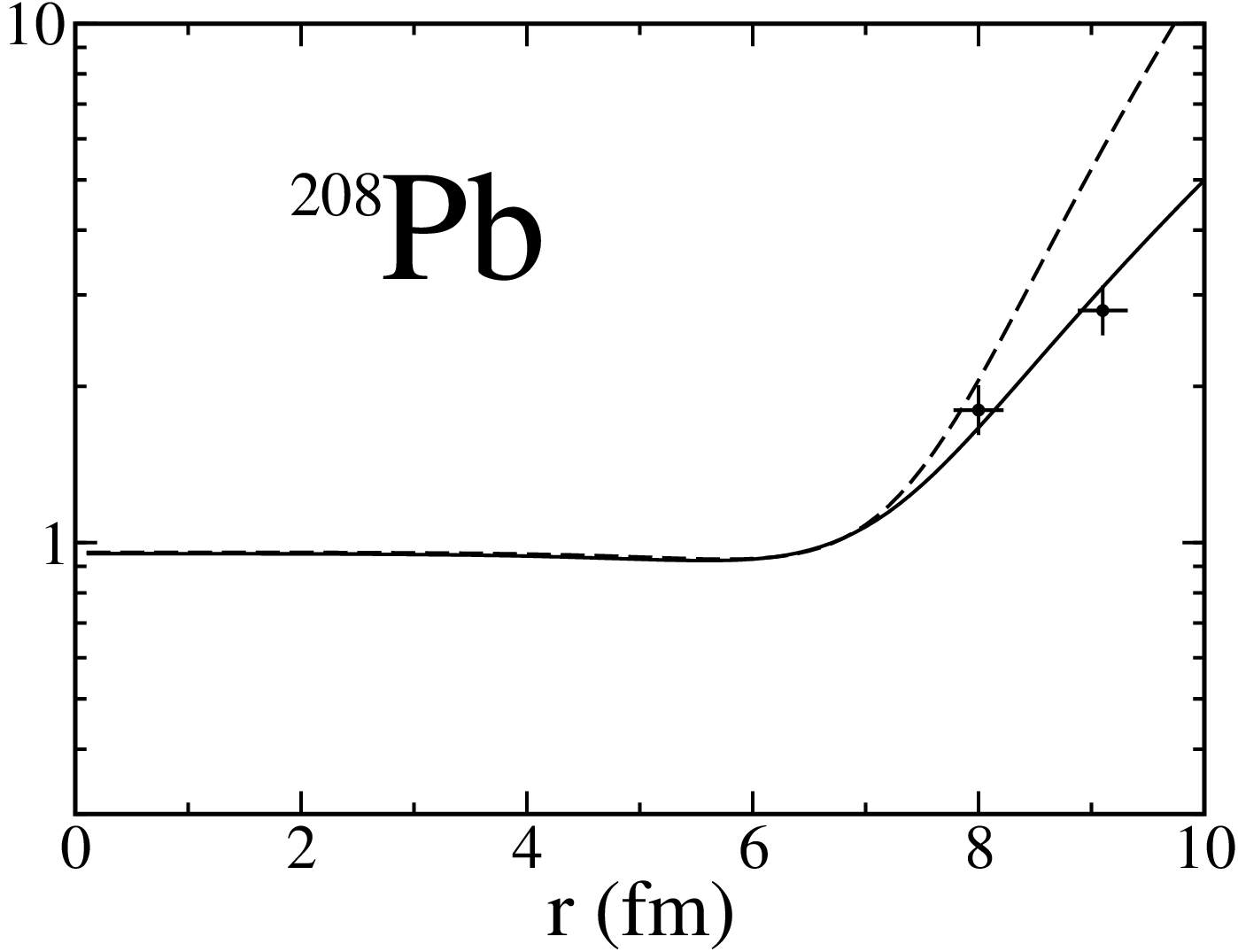}
\end{center}
\caption{The halo factor in $^{124}$Sn and $^{208}$Pb as a function of the distance from the center
of the nucleus calculated using the Eq. (\ref{fhalo}). The values of the halo factor deduced from
experiment are marked by crosses \cite{trja01}.}
\label{fig6}
\end{figure}

In \figurename~\ref{fig6} we show the calculations of the theoretical halo factor for two values of the parameter
$\eta$ together with the experimental data \cite{trja01}. As in the previous figures the solid lines correspond
to the calculation with $\eta=4$ and dashed lines correspond to $\eta=2$.
The left panel is the case for tin isotope $^{124}$Sn, and the right panel is for $^{208}$Pb. 
It is seen from \figurename~\ref{fig6} that the calculation with $\eta=4$ describes the experimental 
data better than the one with $\eta=2$.

It should be noted here that for all calculations for nuclei with mass numbers $A$ 
from 90 to 238 the leptodermous parameter $a/R$ does not exceed 0.1.

\section{Conclusions}

We have considered the neutron skin and halo effects for medium and heavy nuclei within the extended
Thomas-Fermi approximation using effective Skyrme-like forces and the direct variational method.
The nucleon densities $\rho _{p}(\mathbf{r})$ and $\rho _{n}(\mathbf{r})$ are generated by the
profile functions which are obtained by the requirement that the energy of the nucleus should
be stationary with respect to variations of these profiles.

Using the leptodermous properties of the profile nucleon densities $\rho_{p}(\mathbf{r})$ and
$\rho _{n}(\mathbf{r})$, we have obtained the analytical expression for the isovector shift of
the rms radii $\Delta r_{np}$ as a superposition of the two terms.  The first one, $\Delta r^{skin}_{np}$,
describes the neutron "skin-type" distribution and the second one, $\Delta r^{halo}_{np}$, describes
the neutron "halo-type" distribution.

Numerical calculations show that the partial contributions to the neutron-skin thickness from
the skin effect, $\Delta r_{np}^\mathrm{skin}/\Delta r_{np}$, and the halo effect,
$\Delta r_{np}^\mathrm{halo}/\Delta r_{np}$, depend on the parameter $\eta$.
With the inscrease of $\eta$ from $2$ to $\eta=4$ the partial contributions to the neutron-skin
thickness from the skin effect decreases from 100\% to about of 50\% and the halo component increses
from zero to about of 50\%.

The calculated isovector shift of the rms radii $\Delta r_{np}$ and halo factor
$f_\mathrm{theor}^\mathrm{halo}(r)$ give satisfactory description for the corresponding experimental 
data at $\eta=4$. In this case we have superposition of the neutron skin and halo effects with
approximately equal contributions.

{\bf Acknowledgments} 

We would like to thank Prof. V.M. Kolomietz for stimulating our interest to the subject of the 
article and for useful discussions.

\bigskip

\setcounter{equation}{0} \renewcommand{\theequation}{A\arabic{equation}} 
\appendix
\appendix{\bf APPENDIX A}

In this Appendix, we will consider the \textrm{rms} radius for nucleons as
\begin{equation}
\sqrt{\left\langle r^2_q \right\rangle}
=\sqrt{\frac{\int d\mathbf{r}\,r^2\,\rho_q(r)}{\int d\mathbf{r}\,\,\rho_q(r)}}
=\sqrt{\frac{I_{4,q}}{I_{2,q}}},
\label{arms}
\end{equation}
where
\begin{equation}
I_{n,q}=\int_0^\infty dr\,r^n\,\rho_q(r).
\end{equation}
Whithin the leptodermous approximation $a_q/R_q<<1$ one has
\begin{eqnarray}
I_{2,q}&\simeq& \frac{R_q^3}{3}\left(1+3\kappa_0(\xi)\frac{a_q}{R_q}+6\kappa_1(\xi)\left(\frac{a_q}{R_q}\right)^2 
+3\kappa_2(\xi)\left(\frac{a_q}{R_q}\right)^3\right), 
\label{i2} \\
I_{4,q}&\simeq& \frac{R_q^5}{5}\left(1+5\kappa_0(\xi)\frac{a_q}{R_q}+20\kappa_1(\xi)\left(\frac{a_q}{R_q}\right)^2 
+30\kappa_2(\xi)\left(\frac{a_q}{R_q}\right)^3+20\kappa_3(\xi)\left(\frac{a_q}{R_q}\right)^4 \right. \nonumber \\
&   &+\left. 5\kappa_4(\xi)\left(\frac{a_q}{R_q}\right)^5 \right),
\label{i4}
\end{eqnarray}
where $\kappa_j(\xi)$ are the generalized Fermi integrals derived in Ref. \cite{kosa08} 
\begin{equation}
\kappa_j(\xi )=\int_0^\infty {dx\,x^j\left[ (1+e^x)^{-\xi}-(-1)^j\left(1-(1+e^{-x})^{-\xi}\right)\right]}.
\label{ki}
\end{equation}
Inserting Eqs. (\ref{i2}) and (\ref{i4}) into the Eq. (\ref{arms}) one gets
\begin{eqnarray}
\sqrt{\left\langle r^2 \right\rangle_q}\simeq
\sqrt{\frac{3}{5}} R_q \left\{1+\kappa_0(\xi)\frac{a_q}{R_q}-\frac{7}{2}\left(\kappa_0^2(\xi)
-2\kappa_1(\xi)\right)\left(\frac{a_q}{R_q}\right)^2 \right. \nonumber \\
+\left. \frac{1}{6}\left(75\kappa_0^3(\xi)-204\kappa_0(\xi)\kappa_1(\xi)+81\kappa_2(\xi)\right)
\left(\frac{a_q}{R_q}\right)^3 \right\} \label{a6}
\end{eqnarray}
From Eq. (\ref{a6}) the isovector shift of the rms radii $\Delta r_{np}$ reads
\begin{eqnarray}
\Delta r_{np}&=&\sqrt{\left\langle r^2_n \right\rangle}-\sqrt{\left\langle r^2_p \right\rangle} \nonumber \\
&\simeq &\sqrt{\frac{3}{5}} \left\{R_n-R_p+\kappa_0(\xi)(a_n-a_p)
-\frac{7}{2}\left(\kappa_0^2(\xi)-2\kappa_1(\xi)\right)\left(\frac{a_n^2}{R_n}-\frac{a_p^2}{R_p}\right) \right. \nonumber \\
& &+\left. \frac{1}{6}\left(75\kappa_0^3(\xi)-204\kappa_0(\xi)\kappa_1(\xi)+81\kappa_2(\xi)\right)
\left(\frac{a_n^3}{R_n^2}-\frac{a_p^3}{R_p^2}\right) \right\} \label{a7}
\end{eqnarray}
One can rewrite last terms of Eq. (\ref{a7}) using the relations
\begin{equation}
\frac{a_n^2}{R_n}-\frac{a_p^2}{R_p} \simeq
-\left(\frac{a_n+a_p}{R_n+R_p}\right)^2 \left(R_n-R_p\right)+2\frac{a_n+a_p}{R_n+R_p}(a_n-a_p),
\end{equation}
\begin{equation}
\frac{a_n^3}{R_n^2}-\frac{a_p^3}{R_p^2} \simeq
-2\left(\frac{a_n+a_p}{R_n+R_p}\right)^3 \left(R_n-R_p\right)+3\left(\frac{a_n+a_p}{R_n+R_p}\right)^2(a_n-a_p)
\end{equation}
Finally, we have
\begin{eqnarray}
\Delta r_{np} &\simeq &\sqrt{\frac{3}{5}} 
\left\{1+\frac{7}{2}\left(\kappa_0^2(\xi)-2\kappa_1(\xi)\right)\left(\frac{a_n+a_p}{R_n+R_p}\right)^2 \right. \nonumber \\
& &- \left. \frac{1}{3}\left(75\kappa_0^3(\xi)-204\kappa_0(\xi)\kappa_1(\xi)+81\kappa_2(\xi)\right)\left(\frac{a_n+a_p}{R_n+R_p}\right)^3
\right\}\left(R_n-R_p\right) \\
&+& \sqrt{\frac{3}{5}} \left\{\kappa_0(\xi)-7\left(\kappa_0^2(\xi)-2\kappa_1(\xi)\right) \frac{a_n+a_p}{R_n+R_p}\right. \nonumber \\
& &+ \left. \frac{1}{2}\left(75\kappa_0^3(\xi)-204\kappa_0(\xi)\kappa_1(\xi)+81\kappa_2(\xi)\right)
\left(\frac{a_n+a_p}{R_n+R_p}\right)^2 \right\}(a_n-a_p)
\end{eqnarray}

\end{document}